\documentclass[showpacs,twocolumn,preprintnumbers,amsmath,amssymb]{revtex4}
\usepackage{amsmath,amsfonts,latexsym,amssymb,graphicx,graphics,epsfig,subfigure,color,makeidx}
\usepackage{xcolor,diagbox}
\usepackage{multirow}
\usepackage[colorlinks,linkcolor=blue,anchorcolor=blue,citecolor=green,urlcolor=blue]{hyperref}
\newcommand {\nn}{\nonumber}

\begin{document}

\title{Effective action of a self-interacting scalar field on brane}
\author{Cheng-Chen Li$^{a}$$^{c}$}
\author{Zheng-Quan Cui$^{a}$$^{c}$}
\author{Tao-Tao Sui$^{a}$$^{c}$}
\author{Yu-Xiao Liu$^{a}$$^{b}$$^{c}$\footnote{liuyx@lzu.edu.cn, corresponding author}}

\affiliation{$^{a}$Institute of Theoretical Physics $\&$ Research Center of Gravitation, Lanzhou University, Lanzhou 730000, China\\
  $^{b}$Key Laboratory for Magnetism and Magnetic of the Ministry of Education, Lanzhou University, Lanzhou 730000, China\\
  $^{c}$Joint Research Center for Physics, Lanzhou University and Qinghai Normal University, Lanzhou 730000, China\\}

\begin{abstract}
In extra dimensional theories, the four-dimensional
field theory is reduced from a fundamental field theory in the bulk spacetime by integrating the extra dimensional part. In this paper we investigate the effective action of a self-interacting scalar field on a brane in the  five-dimensional thick braneworld scenario. We consider two typical thick brane solutions and obtain the P\"{o}schl-Teller and harmonic potentials of the Kaluza-Klein (KK) modes, respectively. The analytical mass spectra and wave functions along extra dimension of the KK modes are obtained. Further, the effective coupling constant between different KK particles, cross section, and decay rate for some processes of the KK particles are related to the fundamental coupling in five dimensions and the new physics energy scale. Some interesting properties of these interactions are found with these calculations. The KK particles with higher mode have longer lifetime, and they almost do not interact with ordinary matter on the brane if their mode numbers are large enough. Thus, these KK particles with higher modes might be a candidate of dark matter.
\end{abstract}

\maketitle

\section{Introduction}

After it was proposed in the 1920s, Kaluza-Klein (KK) theory \cite{Kaluza1921,Klein1926} was discussed
repeatedly over the past 100 years. The essential idea of this theory
is that the observable physical phenomena of our four-dimensional world can be reduced
from a five-dimensional fundamental theory. According to
string theory, the existence of extra dimensions is inevitable. Various
models containing extra dimensions were proposed and discussed seriously
as effective theories at certain energy level. In the braneworld
scenario \cite{Rubakov1983a,ADD1998,RS1,RS2},
our world is described as a four-dimensional brane embedded in a higher-dimensional spacetime called bulk and extra dimensions
are suggested to be extended far more larger than they were thought before \cite{ADD1998}, or
even infinite \cite{Rubakov1983a,RS2}.

The development of extra dimension theories gives new insight into fundamental physics.
Nevertheless, the reasons that introduce the concept of extra dimensions
are quite different. In KK theory, the fifth dimension was proposed
in order to unify the four-dimensional gravity and  electromagnetic force by regarding the electromagnetic
field as the fifth component of the metric. In Arkani-Hamed-Dimopoulos-Dvali (ADD) model \cite{ADD1998} and Randall-Sundrum (RS)-1 model \cite{RS1}, the existence of extra dimensions provides an alternative mechanism to resolve the gauge hierarchy problem in
particle physics. On the other hand, some efforts have been made to explore the possibility of infinite extra dimensions, such as RS-2 model \cite{RS2} and domain wall model \cite{Rubakov1983a}. Combining the above two models \cite{Rubakov1983a,RS2}, one can obtain the model of a brane with thickness (called thick brane)
in a five-dimensional curve spacetime with an infinite extra dimensions. Such brane is called thick brane. In contrast, the branes in the RS-1 and RS-2 models are called thin branes since they are a pure four-dimensional hypersurface without thickness.

We may regard thin brane as an approximation to thick brane if
the thickness of a brane can be neglected. However, there is a significant difference
between thin brane and thick brane. In a thick brane model, there
are no special dimensions as ``extra dimensions'' since the four-dimensional fields in the Standard Model, which are described by zero modes of some higher-dimensional fields, are lived in the bulk. However, those zero modes are localized around the thick brane and they can not propagate along extra dimensions. For a thin brane model, there are some very special extra dimensions and all the matter fields are confined on a hypersurface, i.e., the brane. Both thin and thick branes have their respective advantages. For example, thin brane resolves the hierarchy problem, while thick brane has its application in holographical QCD.

For all kinds of extra dimension theories, an important question
is why we stay in a four-dimensional world and do not observe these extra dimensions. In the original KK theory it was explained by compacting the extra dimension into a tiny circle with size of Planck
scale so that we can not find it in current experiments even if we actually occupy the whole
volume of the five dimensions. In ADD model and RS-1/RS-2 model, it is
a prior hypothesis that all matter fields are confined on a four-dimensional
hypersurface embedded in five-dimensional spacetime. In the thick
brane model which we mainly concern in this paper, all fields live
in the bulk and the extra dimension extends infinitely, but the matter fields
is trapped in a very narrow region along the extra
dimension. This mechanism is called as localization. There are a lot of references for the thick brane models and localization of gravity and matter fields  \cite{DeWolfe2000,Csaki2000,Kobayashi2002,Wang2002,Liu0907.4424,LuHong2012,Bazeia2015,SouzaDutra2015,LiLiu2017,Mazani2020,Wan:2020smy,Fu:2021rgu,Fu:2022hot,Rosa:2022fhl}, see Refs.~\cite{Dzhunushaliev2010,Rizzo2010,Maartens2010,Raychaudhuri2016,Liu1707.08541,Liu2205.04754} for reviews.

Another principal problem need to be considered is how to recover ordinary
four-dimensional theory such as electroweak gauge theory and
general relativity from the underlying five-dimensional theory. It
leads to the idea that ordinary four-dimensional physics
corresponds to some five-dimensional physics. From the very popular
AdS/CFT perspective, our ordinary four-dimensional
physics, as a field theory living on the boundary of a five-dimensional
spacetime, is totally equivalent to a gravitational theory in the
 bulk. On the other hand, in KK theory, domain
wall model and brane model, it is accomplished by reducing
the high-dimensional fundamental theory to the four-dimensional effective one, which is realized by integrating the high-dimensional action over extra dimensions.
% to obtain the four-dimensional one.
%Although braneworld were considered deeply related to AdS/CFT, it is not a manifestly holographical duality, since information were lost after integral.

Most of the works about localization of a scalar field on a thick brane focus on free field and few works discuss the self-interaction of a scalar field in the bulk as well as its effective theory on a thick brane.
In this paper we will investigate the effective action of a self-interacting scalar field on a thick brane. We first assume a five-dimensional fundamental action of a self-interacting scalar field. Then we employ the KK reduction procedure to derive the four-dimensional effective actions of the scalar KK modes. The effective action should be coincident with the action in current four-dimensional theory. Furthermore, the five-dimensional interaction will bring new four-dimensional particles and new interaction terms between them. This will make some significant prophecies on particle collision at higher energy level.

The organization of this paper is as follows. In Sec.~\ref{sec_2}, we give a general formulation for the KK reduction of a five-dimensional action of a self-interacting scalar field. In Sec.~\ref{sec_3}, we consider two brane solutions and obtain two kinds of typical potentials which demonstrate some significant properties of the four-dimensional effective action. Finally, in Sec.~\ref{sec_4} we devote to the conclusions and discussions.

\section{KK reduction of action} \label{sec_2}

Let us first consider a free massless scalar field in five-dimensional spacetime:
\begin{equation}
S=\int\sqrt{-g}\,d^{5}x\left(-\frac{1}{2} g^{MN} \partial_{M}\phi\partial_{N}\phi\right). \label{ActionS5}
\end{equation}
The five-dimensional metric is proposed as
\begin{eqnarray}
ds^2 = g_{MN} dx^M dx^N
     = e^{2A(z)} \left( \tilde{g}_{\mu\nu}(x^\lambda) dx^\mu dx^\nu +dz^2 \right), \label{5Dmetric}
\end{eqnarray}
where $M,N$ and $\mu,\nu$ are the five-dimensional and four-dimensional coordinate indices, $z=x^{5}$ is the extra dimensional coordinate, and $\tilde{g}_{\mu\nu}(x^\lambda)$ is the reduced four-dimensional metric at any fixed position of the extra dimensional coordinate $z=z_0$ by a constant factor $e^{2A(z_0)}$. Usually, we can set $e^{2A(0)}=1$ by using the degree of freedom of the coordinate transformation $x^\mu\rightarrow \bar{x}^{\mu} = e^{A(0)} x^\mu$.

In order to reduce the five-dimensional fundamental action (\ref{ActionS5}) to
a four-dimensional effective one, we separate the variables of extra
dimension from ordinary four dimensions
\begin{equation}
\phi (x^{\mu},z) =\underset{n}{\sum}\varphi_{n}(x^{\mu})f_{n}(z), \label{KKDecomposition}
\end{equation}
which is called the KK decomposition. Substituting the above decompsition into the action (\ref{ActionS5}), we
have
\begin{eqnarray}
S&=&-\frac{1}{2}\int \sqrt{-\tilde{g}} d^{4}x\int e^{5A(z)}\,dz
\underset{n,m}{\sum}
          \big(f_{n}f_{m}  {\partial}_{\mu}\varphi^{n}\partial_{\nu}\varphi_{m}\\ \nn
&+&\varphi_{n}\varphi_{m}\partial^{5}f_{n}\partial_{5}f_{m}\big).
\end{eqnarray}

The field equation corresponding to the action (\ref{ActionS5}) is the five-dimensional
Klein-Gordon equation:
\begin{equation}
g^{MN}\nabla_{M}\nabla_{N}\phi=0. \label{5DKGEq}
\end{equation}
By virtue of the KK decomposition (\ref{KKDecomposition}), the above Klein-Gordon equation (\ref{5DKGEq})  can be converted to two
equations. One is the familiar four-dimensional Klein-Gordon equation
of four-dimensional modes $\varphi_{n}(x^{\mu})$ and another is an eigenvalue equation of the extra-dimensional part
$f_{n}(z)$ :
\begin{eqnarray}
({\square}^{(4)}-\zeta_{n})\varphi_{n}(x^{\mu}) &=& 0,\\
(\partial_{z}^{2}+3(\partial_{z}A)\partial_{z})f_{n}(z) &=& \zeta_{n}f_{n}(z),\label{eq:f}
\end{eqnarray}
where ${\square}^{(4)}=\tilde{g}^{\mu\nu}\nabla_{\mu}\nabla_{\nu}$.
With the redefinition of the field
\begin{equation}
\chi_{n}(z)=e^{\frac{3}{2}A}f_{n}(z),\label{redefine_chi}
\end{equation}
Eq. (\ref{eq:f}) becomes a one-dimensional Schr\"{o}dinger-like
equation
\begin{equation}
\left[-\partial_{z}^{2}+V(z)\right]\chi_{n}(z)=\zeta_{n}\chi_{n}(z), \label{SchrodingerEq}
\end{equation}
where the effective potential is
\begin{equation}
V(z)=\frac{3}{2}\partial_{z}^{2}A+\frac{9}{4}(\partial_{z}A)^{2}.  \label{potential_V}
\end{equation}

If the eigenstate $\chi_{n}(z)$ of the Schr\"{o}dinger-like equation
is a normalizable bound state, the corresponding
KK mode is localized on a brane, in the sense that the energy density
of the field actually distributes in a finite region of the extra
dimension and one can obtain the four-dimensional action of the KK mode $\varphi_{n}(x^{\mu})$. To this end, we require the normalization condition as well as in quantum mechanics:
\begin{equation}
\int\chi_{n}\chi_{m}\,dz=\delta_{nm},  \label{normalizationCondition1}
\end{equation}
or equivalently
\begin{equation}
\int e^{3A(z)}f_{n}f_{m}\,dz=\delta_{nm}, \label{normalizationCondition2}
\end{equation}
which could be inferred from Eq. (\ref{SchrodingerEq}) that
\begin{equation}
\int e^{3A(z)}\partial_{z}f_{n}\partial_{z}f_{m}\,dz=\delta_{nm}\sqrt{\zeta_{n}\zeta_{m}},
 \label{normalizationCondition3}
\end{equation}
where $m,n$ are not summarized here. The ground state with $\zeta_0=0$ is called the zero
mode and is interpreted as the ordinary four-dimensional scalar field that we have
observed on the brane.

By virtue of the normalized conditions (\ref{normalizationCondition2}) and (\ref{normalizationCondition3}), the action (\ref{ActionS5})
is reduced to
\begin{equation}
S_{\text{eff}}=\int \sqrt{-\tilde{g}} d^{4}x
  \left[-\frac{1}{2}\underset{n}{\sum}
        \left(
            \partial^{\mu}\varphi_{n}\partial_{\mu}\varphi_{n}+\zeta_{n}\varphi_{n}^{2}
        \right)
  \right],
\end{equation}
where ${\partial}^{\mu} \equiv \tilde{g}^{\mu\nu} \partial_{\nu}$ and
$\zeta_{n}$ is the eigenvalue of Eq. (\ref{eq:f}) and it can be
proven to be nonnegative. The result is interpreted as a family of
four-dimensional scalar fields with different masses and $\sqrt{\zeta_{n}}$
is called induced mass originating from extra dimension.
It can be seen in another way that a massless particle in five-dimensional
spacetime satisfies the following relation
\begin{equation}
\tilde{g}^{\mu\nu} p_{\mu}p_{\nu} = - (p_{5})^2.
\end{equation}
In case of a nonzero momentum along the extra dimension, the four-dimensional particle on the brane has an effective mass $m=|p_{5}|$.

It should be underlined that in our formulation, by virtue of the
KK decomposition of the scalar field, the equation for the extra dimensional
part $f_{n}(z)$ of the scalar field decouples from the equation of
the four-dimensional part $\varphi_{n}(x^{\mu})$. This guarantees
we can study physics on the brane, otherwise we will be blind to
a dynamical variable and we never have complete dynamical equations.
The approach of the effective action on the brane accords with such
an assertion that {all observers have a right to describe physics
using an effective theory based only on the variables they can access}.
In particle physics we have met this concept in the renormalization
group theory. To describe particle interactions at 10 GeV in the lab,
we do not need to know what happens at $10^{14}$ GeV. We have also
seen that this concept is demonstrated in the cosmic censorship that a
horizon protects our predictive ability on the basis of general relativity
from a singularity. In the braneworld scenario here, if observers
are confined on brane, they should still be able to study physics
using only the variables accessible to them without having to know
what happens outside of the brane.

So far the scalar field is free. We would like to include interaction
by adding a perturbative interaction potential $U(\phi)$ in the action (\ref{ActionS5}):
\begin{equation}
S_{\text{int}}=-\int \sqrt{-\tilde{g}}\,  d^{4}x\int dz\,e^{5A(z)}\,U(\phi).
\end{equation}
We expect that the following four-dimensional effective interaction term can be given by integrating the above interaction term over the extra dimension:
\begin{align}
S_{\text{int}}^{(4)} & =-\int\sqrt{-\tilde{g}}\,  d^{4}x\,U^{(4)}(\varphi(x^{\mu})),
\end{align}
where
\begin{equation}
U^{(4)}(\varphi(x^{\mu}))=\int dz\,e^{5A(z)}\,U\left(\phi\left(x^{\mu},z\right)\right).
\end{equation}
It is just an integral projection from five-dimensional spacetime
to our four-dimensional one.
%As long as the integral $\int dz\, e^{5A(z)}\,U\left(\phi\left(x^{\mu},z\right)\right)$
%is calculated, we would obtain the four-dimensional effective
%interaction term
%\begin{align}
%S_{\text{int}}^{(4)} & =-\int d^{4}x\,U^{(4)}(\varphi(x^{\mu})).
%\end{align}
Comparing to the ordinary four-dimensional action which only involves
the zero mode $\varphi_{0}$
\begin{equation}
S_{0}^{(4)}=\int  \sqrt{-\tilde{g}}\, d^{4}x
  \left[-\frac{1}{2}
        \left(\partial^{\mu}\varphi_{0}\partial_{\mu}\varphi_{0}
              +\zeta_{0}\varphi_{0}^{2}
        \right)
        -U_{0}^{(4)}(\varphi_{0})
  \right],
\end{equation}
it can be seen that
\begin{eqnarray}
S^{(4)}&=&S_{0}^{(4)} +
  \int \sqrt{-\tilde{g}}\, d^{4}x
   \Big[-\frac{1}{2}\underset{n\ge 1}{\sum}
\big(\partial^{\mu}\varphi_{n}\partial_{\mu}\varphi_{n}\\ \nn
        &+&\zeta_{n}\varphi_{n}^{2}
        \big)
       -\left(U^{(4)}(\varphi)-U_{0}^{(4)}(\varphi_{0})
        \right)
  \Big]. \label{S4}
\end{eqnarray}
The second term in the four-dimensional action (\ref{S4}) is new and totally originates from
the extra dimension. It predicts not only new massive particles but
also new interactions: $S_{\text{int}}^{(4)}$ contains more terms other
than the ordinary four-dimensional interaction $S_{0\: \text{int}}^{(4)}=-\int d^{4}x\,U_{0}^{(4)}$.
It contains all possible interactions of various KK modes $\varphi_{n}$.

\begin{figure*}[htbp]
    \makebox[\textwidth][c]{
  \includegraphics[width=0.2\textwidth]{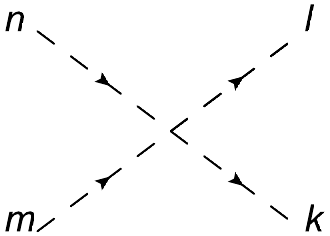}
  }
    \caption{The Feynman diagram for (\ref{Sint_klmn}).}
    \label{fig:FeynmanDiagram_nlmk}
\end{figure*}

In this paper we will investigate the interaction of a quartic form
\begin{equation}
U=\lambda\phi^{4}=\lambda\bigg[\underset{n}{\sum}\varphi_{n}(x^{\mu})f_{n}(z)\bigg]^{4}.
\end{equation}
It can be expressed in a four-dimensional effective action as
\begin{equation}
S_{\text{int}}^{(4)} =-\int  \sqrt{-\tilde{g}}\, d^{4}x\,\lambda\underset{klmn}{\sum}\gamma_{klmn}\varphi_{k}\varphi_{l}\varphi_{m}\varphi_{n}, \label{Sint_klmn}
\end{equation}
where
\begin{equation}
\gamma_{klmn}=\int e^{5A(z)}\,dzf_{k}f_{l}f_{m}f_{n}.  \label{coupling_gamma_klmn}
\end{equation}
As we known in general, $f_{k},f_{l},f_{m},f_{n}$ are four different
or same KK modes. From the four-dimensional viewpoint, there are a
family of scalar KK particles with different masses $\sqrt{\zeta_{n}}$.
They interact with each other with the form
(\ref{Sint_klmn}). The coupling constant $\lambda\gamma_{klmn}$ is exactly the
scattering amplitude on tree level with the Feynman diagram showed
in Fig. \ref{fig:FeynmanDiagram_nlmk}. We will calculate them in the next section.

Furthermore, there is a very interesting difference between the cases
of interactive and free scalar fields. It is convinced that we cannot
measure the distribution of the field on extra dimension since we
are restricted on the brane. What we could measure on brane is just
the mass spectrum of the KK particles. So a problem arises: can we
discover the metric of five-dimensional spacetime while the observers
and all our observations are confined on a four-dimensional brane?

This question is equivalent to the quantum mechanics problem that, can
we reconstruct the shape of a potential $V(z)$ from its eigenvalue spectrum
$\{E_{n}\}$? Unfortunately, the answer is negative. For example,
the Morse potential
\begin{equation}
V(z)=C_{1}^{2}+e^{-2z}-(2C_{1}+1)e^{-z}
\end{equation}
and the Scarf (hyperbolic) potential
\begin{eqnarray}
 V(z)&=&-[C_{1}(C_{1}+1)-C_{2}^{2}]\text{sech}^2 z\\ \nn
      &+&(2C_{1}+1)C_{2}\tanh z ~\text{sech}z
      +C_{1}^{2} % ,\quad-\infty<z<\infty,
\end{eqnarray}
%and the P$\ddot{\mathrm{o}}$schl-Teller (PT) (hyperbolic) potential
%\begin{equation}
%V(z)=[C_{1}(C_{1}+1)+C_{2}^{2}]\cosh^{2}z+C_{1}^{2}
%    -(2 C_{1} + 1) C_{2} \cosh z~ \coth z, \quad 0<z<\infty,~C_{1}<C_{2}.
%\end{equation}
have exactly the same eigenenergies:
\begin{equation}
E_{n}=C_{1}^{2}-(C_{1}-n)^{2},
\end{equation}
even though the potentials and eigenfuctions are quite distinct.
Here $C_{1}$ and $C_{2}$ are constants and $-\infty<z<\infty$.
In fact, we could learn
form supersymmetric quantum mechanics that there are a large number
of potentials that have the same eigenenergies and are related with each
other by the so called isospectral deformations. We do not strive
for the details here. There seems to be an inevitable conclusion:
the five-dimensional metric can not be known if we just stay on the
brane. Nevertheless,
interactions of various KK modes will change this situation.

With a fundamental interaction, say, $\phi^{4}$, we could measure not only the
mass spectrum $\{m_{n}\}$ of the KK particles, but also the coupling
constants $\{\lambda\gamma_{klmn}\}$, which would distinguish the
potentials corresponding to the same eigenvalues. The point is that the
coupling constants in Eq.~(\ref{coupling_gamma_klmn}) involve eigenfunctions $\{f_{n}\}$,
which are distinct for different potentials. So we have seen the significance
of interaction: the effective coupling constants reveal the structure
of five-dimensional spacetime which cannot be inferred by the free
fields on the brane. On the other hand, the interaction provides channels that KK particles
transform into each other. The massive KK modes may decay to the zero mode, while they can also be produced from collision of the zero mode particles.

\section{Properties of the effective action}\label{sec_3}

We have seen that the effective potential of the KK modes is only determined by the warp factor $A(z)$ in the bulk metric (\ref{5Dmetric}):
\begin{equation}
V(z)=\frac{3}{2}\partial_{z}^{2}A+\frac{9}{4}(\partial_{z}A)^{2}, \label{potentialVz}
\end{equation}
which determines the KK modes $f_{n}$ as well as the effective coupling constants $\lambda\gamma_{klmn}$.
In this paper, we only use some explicit forms of the warp factor $A(z)$ instead of going into the details of solving the field equations for the background spacetime, which depend on the chosen gravitational theory and the background fields generating the thick brane.
We will discuss two kinds of typical potentials which demonstrate some significant properties of the
four-dimensional effective action.

\begin{figure*}[htbp]
    \makebox[\textwidth][c]{%\includegraphics[width=50mm]{PT}
    \includegraphics[width=0.3\textwidth]{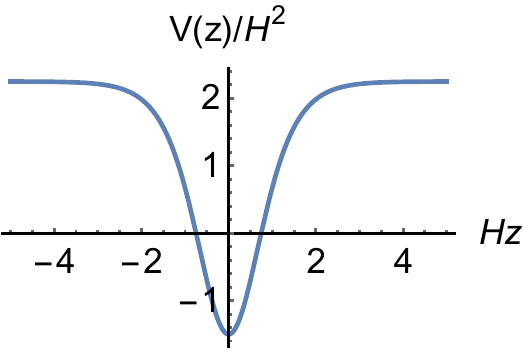}
    }
    \caption{The shape of the potential (\ref{PTpotential}).}
    \label{fig:PTpotential}
\end{figure*}

{In Refs.~\cite{Herrera-Aguilar:2010ehj,Guo2013}, the authors presented  the de Sitter braneworld model, in which an induced 3-brane with spatially flat cosmological background is considered. The action and the five-dimensional metric for this braneworld model are given by
\begin{eqnarray}
   S\!\!\!&=&\!\!\!\int d^5x\sqrt{-g} \frac{1}{2\kappa^2_5}
       ( R -2\Lambda_5),\\
ds^2\!\!\!&=&\!\!\!e^{2A(z)}\Big[-dt^2+a(t)^2(dx_1^2+dx_2^2+dx_3^2)+dz^2\Big],~~~
\end{eqnarray}
where $\kappa^2_5=1/M_\ast^3$ with $M_\ast$ being the five-dimensional fundamental scale, $\Lambda_5$ is the bulk cosmological constant, and $a(t)$ is the scale factor of the brane.
The brane solution is~\cite{Herrera-Aguilar:2010ehj,Guo2013}
\begin{eqnarray}
 A(z)&=&\ln\left(\frac{H}{b}\mathrm{sech}(bz)\right),\\
 a(t)&=&e^{Ht},\\
 \Lambda_5&=& 6 b^2,
\end{eqnarray}
where $1/b$ parameterizes the thickness of the brane, the parameter $H$ is the Hubble parameter. The relation between the effective four-dimensional cosmological constant on the brane and the Hubble parameter $H$ is $\Lambda_4 = 3H^2$. In this paper, we consider the special case of $b=H$ for simplicity. For the de Sitter branworld model, the corresponding potential is the P$\ddot{\mathrm{o}}$schl-Teller (PT) potential}
\begin{equation}
V(z)=\frac{3}{4}H^{2}\left[3-5\text{sech}^{2}(Hz)\right], \label{PTpotential}
\end{equation}
which is shown in Fig. \ref{fig:PTpotential}.
For this potential, we
have two bound states. The first one is the ground state,
i.e., the zero mode
\begin{equation}
\chi_{0}(z)=e^{\frac{3}{2}A}f_{0}(z)=\sqrt{\frac{2H}{\pi}}\mathrm{sech^{\frac{3}{2}}}(Hz)
\end{equation}
with mass
\begin{equation}
m_{0}=\sqrt{\zeta_{0}}=0.
\end{equation}
The second one is the first massive KK mode
\begin{equation}
\chi_{1}(z)=e^{\frac{3}{2}A}f_{1}(z)=\sqrt{\frac{2H}{\pi}}\mathrm{sech^{\frac{3}{2}}}(Hz)\sinh(Hz)
\end{equation}
with mass
\begin{equation}
m_{1}=\sqrt{\zeta_{1}}=\sqrt{2}H.
\end{equation}
Here, we can see that the parameter $H$ can be viewed as the new physics energy scale since it determines the mass of the new particle (the first massive KK particle) beyond the Standard Model.
Obviously the masses are determined by the warp factor $A(z)$ in
the bulk metric. The two bound states are in fact two kinds
of KK particles on the brane.

\begin{figure*}[htbp]
    \makebox[\textwidth][c]{
  \includegraphics[width=0.2\textwidth]{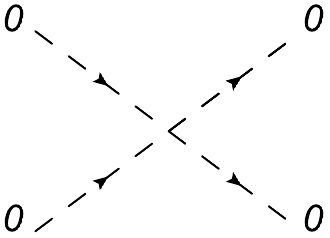}~~~~~~~
  \includegraphics[width=0.2\textwidth]{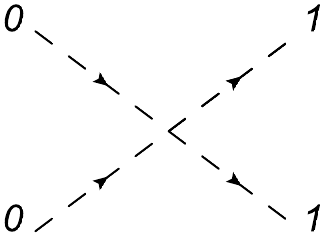}~~~~~~~
  \includegraphics[width=0.2\textwidth]{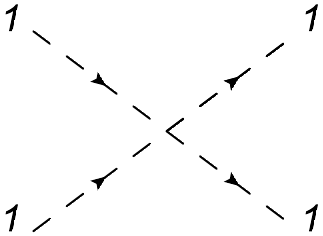}}
    \caption{{The Feynman diagrams for (\ref{0000})-(\ref{1111}).}}
    \label{fig:FeynmanDiagrams_aabb}
\end{figure*}

The two KK modes interact with each other via the four-dimensional
effective interactions from the fundamental scalar potential $\phi^{4}$. The effective coupling constants
are calculated as follows:
\begin{align}
\lambda\gamma{}_{0000} & =\int e^{5A(z)}\,dzf_{0}^{4}=\frac{3H\lambda}{2\pi}, \label{0000}\\
\lambda\gamma{}_{0011} & =\int e^{5A(z)}\,dzf_{0}^{2}f_{1}^{2}=\frac{H\lambda}{2\pi}, \label{0011}\\
\lambda\gamma{}_{1111} & =\int e^{5A(z)}\,dzf_{1}^{4}=\frac{3 H\lambda}{2\pi}, \label{1111}\\
\lambda\gamma{}_{0001} & =\gamma{}_{0111}=0,
\end{align}
which show that there are three kinds of interactions between these two KK modes.
The corresponding Feynman diagrams are shown in Fig. \ref{fig:FeynmanDiagrams_aabb}.

It can be seen that, like the mass spectrum of the KK modes, the effective
coupling constants are determined by the warp factor $A(z)$. Both
of them origin from the extra dimension. However, they associate two
independent parameters $H$ and $\lambda$. We should assume that
the zero mode stands for the four-dimensonal particle which has already
been observed (so that $H\lambda$ can be fixed) and has a self-interacting
$\lambda\gamma{}_{0000}\varphi_{0}^{4}$.

Provided that we make zero mode particles collide with enough energy,
it should be observed in reaction new particles that correspond to
the first excited mode $f_{1}$ (so that $H$ and $\lambda$ can be fixed) as well
as zero mode. On tree level the magnitudes of the reaction are just
the effective coupling constants. The cross sections of the above
two processes are
\begin{eqnarray}
 d\sigma_{00}&=& (2\pi)^{2}\delta^{4}(p_{3}+p_{4}-p_{1}-p_{2})
       \frac{9 H^{2}\lambda^{2}}{\left(-4p_{1}\cdot p_{2}\right)}\\ \nn
       &&*\frac{d^{3}{\vec{p}_{3}}}{(2\pi)^{3}2p_{3}^{0}}
       \frac{d^{3}{\vec{p}_{4}}}{(2\pi)^{3}2p_{4}^{0}}, \\
 d\sigma_{11}&=&(2\pi)^{2}\delta^{4}(p_{3}+p_{4}-p_{1}-p_{2})
       \frac{\pi^{2} H^{2}\lambda^{2}}{\left(-16p_{1}\cdot p_{2}\right)}\\ \nn
     && * \frac{d^{3}{\vec{p}_{3}}} {(2\pi)^{3}2p_{3}^{0}}
       \frac{d^{3}{\vec{p}_{4}}} {(2\pi)^{3}2p_{4}^{0}},
\end{eqnarray}
where $p_{1}$ and $p_{2}$ are 4-momenta of the initial state particles
and $p_{3}$ and $p_{4}$ are 4-momenta of the final state ones. The
total cross sections are the integral of the phase space of the final states
\begin{eqnarray}
\sigma_{00} &=& \frac{1}{4\pi{}^{3}} \frac{9 H^{2}\lambda^{2}}{-8p_{1}\cdot p_{2}},\\
\sigma_{11} &=& \frac{1}{4\pi{}^{3}} \frac{ H^{2}\lambda^{2}}{-8p_{1}\cdot p_{2}}
                \frac{\sqrt{\left({p_{1}^{0}+p_{2}^{0}}\right)^{2}  -8H^{2}} }
                     {\left(p_{1}^{0}+p_{2}^{0}\right)}.
\end{eqnarray}
We see that the branch ratio
\begin{eqnarray}
R(00\rightarrow11)
  &=& \frac{\sigma_{11}}{\sigma_{00}+\sigma_{11}}  \nonumber \\
  &=& 1-\frac{9\left({p_{1}^{0}+p_{2}^{0}}\right)}
             {\sqrt{\left({p_{1}^{0}+p_{2}^{0}}\right)^{2}-8H^{2}}
              +9\left(p_{1}^{0}+p_{2}^{0}\right)}
\end{eqnarray}
is significant only if the initial energy $p_{1}^{0}+p_{2}^{0}$ is
high enough. The above results show that the cross sections and branch ratio are related to the fundamental coupling $\lambda$ in five dimensions and the new physics energy scale $H$.

Note that, the decay of the first excited mode $f_{1}$ into three zero modes $f_{0}$  is prohibited. It implies that there exist
some ``selected rules'' in the interactions between KK particles
just like transitions between states in quantum mechanics. From our
discussion in the end of section 2, in the case of the PT potential,
there must be plenty of $f_{1}$ particles on the brane.

Although the above PT potential illustrates some main features of the effective
action, it is a little simple because it only has two kinds of KK
particles on the brane. There is more interesting content in the case
of harmonic potential since all of its eigenstates are bound
states and we have a series of infinite localized KK particles in
four dimensions. Therefore, we can investigate how the four-dimensional
effective interactions vary while the KK modes go higher.

{To this end, we consider the flat thick brane model generated by a mimetic scalar field with a Lagrange multiplier~\cite{SuiLiu2021}. The action of this theory is given by
\begin{eqnarray}
        S=\int d^5x\sqrt{-g}\left( \frac{R}{2\kappa^2_5}
        + \lambda\big(\partial_M \phi \partial^M \phi-U\big)-V \right),
        \label{action mgb1}
    \end{eqnarray}
where $U=U(\phi)$ and $V=V(\phi)$ are functions of the the scalar field $phi$, and $\lambda$ is a Lagrange multiplier. For simplicity, one can choose the natural unit with $\kappa^2_5=1$.
One of the solutions for the flat thick brane metric (\ref{5Dmetric}) with $\tilde{g}_{\mu\nu}=\eta_{\mu\nu}$ is given by \cite{SuiLiu2021}
\begin{eqnarray}
 A&=&-k^{2}z^{2},  \\
 \phi&=&v \left(\frac{kz}{\sqrt{1+k^2z^2}}\right)^\gamma,
\end{eqnarray}
where the parameter $k$ is the scale parameter which controls the thickness of the brane. Here, we do not list the expressions of $\lambda$, $U$, and $V$. For this brane solution, this warp factor $A(z)$ can generate a harmonic potential in a form of}
\begin{equation}
V(z)=9k^{4}z^{2}-3k^{2}. \label{HarmonicPotential}
\end{equation}
The eigenfunction is
\begin{eqnarray}
\chi_{n}(z)&=& e^{\frac{3}{2}A}f_{n}(z) \nonumber \\
  &=& \left(\frac{3k^{2}}{\pi}\right)^{\frac{1}{4}}
      \frac{1}{\sqrt{2^{n}n!}}e^{-\frac{3}{2}k^{2}z^{2}}H_{n}(\sqrt{3}kz),
\end{eqnarray}
where $H_{n}$ is the Hermitian polynomial
\begin{equation}
H_{n}(\xi)=(-1)^{n}e^{\xi^{2}}\frac{\mathrm{\mathit{\mathrm{d}^{n}}\mathit{e^{-\xi^{2}}}}}{\mathrm{d}\xi^{n}}.
\end{equation}
The four-dimensional induced mass, i.e., the eigenvalue is
\begin{equation}
\zeta_{n}=m_{n}^{2}=6k^{2}n,\quad n=0,1,2,...
\end{equation}
and the effective coupling constant is
\begin{equation}
\lambda\gamma_{klmn}=\lambda\int e^{5A(z)}f_{k}f_{l}f_{m}f_{n}\,dz.\label{klmn}
\end{equation}
Here, we can also see that the parameter $k$ is related to the new physics energy scale.

Notice that, if we add a constant $c$ in the potential (\ref{HarmonicPotential}), the
induced mass will be changed, meanwhile the effective coupling constant
remains unchanged since we have the same configurations for the KK modes. This constant term $c$ may come from the five-dimensional mass term of the scalar field. Thus, we get new induced masses but the same effective coupling constants, i.e., the effective interaction on the brane is independent of the five-dimensional mass of the scalar field.

It is assumed that only the zero mode $\varphi_{0}$ and the interaction
$\lambda\gamma{}_{0000}\varphi_{0}^{4}$ have been observed on the
brane. So they are the ``ordinary'' particles and the four-dimensional
$\varphi^{4}$ interaction, respectively. As demonstrated before,
the excited KK modes emerge and interact with each other when we
increase energy of particle collision to sufficiently high level.

{Note that Eq.~\eqref{klmn} is an integration result. The integrand is too complicated to be figured out for the general modes $(k,l,m,n)$. Hence, we consider the effective coupling constant between KK particles for some lower fixed modes of them. The effective coupling constant of the lowest mode $(0,0,0,0)$ is
\begin{align}\label{eq:lowestcoupling}
  \lambda\gamma{}_{0000} =& \frac{3 k\lambda}{\sqrt{5 \pi }}\,.
\end{align}
This value will be considered as a unit in the following calculations. The rest results will be presented numerically.}

\begin{table*}[!htb]
\begin{center}
\begin{tabular}{|c|c|c|c|c|c|c|c|c|c|c|}
\hline
$n$ & ~~0~~ & 1 & 2 & 3 & 4 & 5 & 10 & 15 & 20 & 30\tabularnewline
\hline
$\frac{\gamma_{nnnn}}{\gamma_{0000}}$ & 1 & 1.08 & 1.43 & 2.10 & 3.28 & 5.37 & 86.7 & 1779 & 4.02$\times10^{4}$ & 2.33$\times10^{7}$\tabularnewline
\hline
\end{tabular}
\caption{The values of $\gamma_{nnnn}$ in unit of $\gamma_{0000}$.}
\label{tab:gamma_nnnn}
\end{center}
\end{table*}

Firstly let us check the ``${nnnn}$'' interaction, i.e., the
interaction in form of $\varphi_n^{4}$ between the same KK modes. We
see that all kinds of KK particles have a $\varphi_{n}^{4}$ interaction.
However, as the quantum number $n$ increasing, the coupling varies form
weak to extraordinary strong, see Tab.~\ref{tab:gamma_nnnn}. Unlike the mass of the $n$-th excited
KK particle which linearly depends on the quantum number $n$, the effective
coupling constant seems to increase as an exponential-like function
of $n$. It may be amazing that, even though the interaction in five-dimensional
spacetime is assumed to be weak, the four-dimensional effective interaction
on brane is not necessarily weak. In fact, it can be very strong.

\begin{table*}[!htb]
\begin{center}
\begin{tabular}{|c|c|c|c|c|c|c|c|c|c|c|c|}
\hline
\diagbox{~~$n$}{~~$m$}  & 0 & 1 & 2 & 3 & 4 & 5 & 6 & 7 & 8 & 9 & 10\tabularnewline
\hline
0 & 1 & 0 & -0.283 & 0 & 0.098 & 0 & -0.036 & 0 & 0.013 & 0 & -0.005\tabularnewline
\hline
1 & 0 & 0.6 & 0 & -0.294 & 0 & 0.131 & 0 & -0.057 & 0 & 0.024 & 0\tabularnewline
\hline
2 & -0.283 & 0 & 0.44 & 0 & -0.277 & 0 & 0.147 & 0 & -0.072 & 0 & 0.034\tabularnewline
\hline
3 & 0 & -0.294 & 0 & 0.36 & 0 & -0.258 & 0 & 0.153 & 0 & -0.083 & 0\tabularnewline
\hline
4 & 0.098 & 0 & -0.277 & 0 & 0.312 & 0 & -0.24 & 0 & 0.155 & 0 & -0.09\tabularnewline
\hline
5 & 0 & 0.131 & 0 & -0.258 & 0 & 0.279 & 0 & -0.225 & 0 & 0.154 & 0\tabularnewline
\hline
6 & -0.036 & 0 & 0.147 & 0 & -0.24 & 0 & 0.255 & 0 & -0.213 & 0 & 0.152\tabularnewline
\hline
7 & 0 & -0.057 & 0 & 0.153 & 0 & -0.225 & 0 & 0.237 & 0 & -0.202 & 0\tabularnewline
\hline
8 & 0.013 & 0 & -0.072 & 0 & 0.155 & 0 & -0.213 & 0 & 0.221 & 0 & -0.192\tabularnewline
\hline
9 & 0 & 0.0241 & 0 & -0.083 & 0 & 0.154 & 0 & -0.202 & 0 & 0.209 & 0\tabularnewline
\hline
10 & -0.005 & 0 & 0.034 & 0 & -0.09 & 0 & 0.152 & 0 & -0.192 & 0 & 0.198\tabularnewline
\hline
\end{tabular}
\caption{The values of $\gamma_{00mn}$ in unit of $\gamma_{0000}$.}
\label{tab:gamma_00mn}
\end{center}
\end{table*}

Next, we consider the ``${00mn}$'' interactions, i.e., the interaction terms containing two zero modes and two massive modes $\varphi_{m},~\varphi_{n}$. Such interactions correspond to the process that two zero mode particles scatter into two KK particles in modes $n$
and $m$. The tree level amplitude is exactly $\lambda\gamma{}_{00mn}$. The result is listed in Tab.~\ref{tab:gamma_00mn}.
Obviously, there is a selected rule: $\gamma_{00mn}$ has a nonzero
value only if $m+n$ is even. This can be easily proven from the parities
of the eigenfunctions. Another significant thing is that the sign
of nonzero $\gamma_{00mn}$ becomes positive and negative alternately,
which means the effective interaction becomes repulsion and attraction
alternately.

\begin{figure*}[htbp]
    \makebox[\textwidth][c]{
     \includegraphics[width=0.4\textwidth]{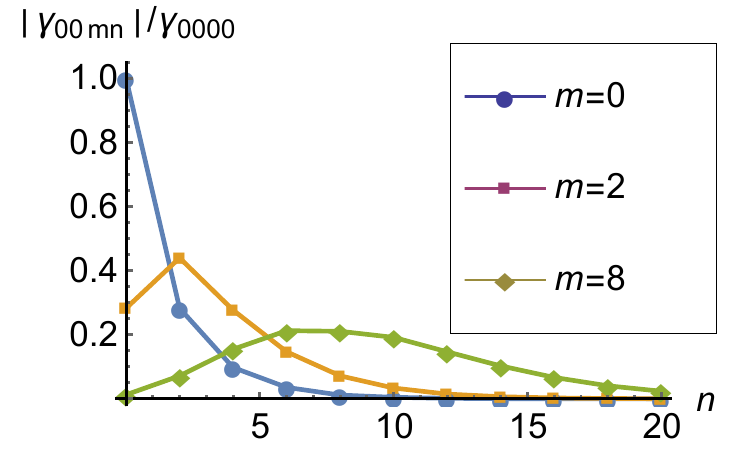}
    }
    \caption{{The relations between nonzero $\left|\gamma_{00mn}\right|$ in unit of $\gamma_{0000}$ versus $n$ for $m=0,2,8$.}}
    \label{fig:relations_00mndata}
\end{figure*}

On the other hand, the relation between the nonvanishing $\gamma_{00mn}$
and $m,n$ is interesting. Figure \ref{fig:relations_00mndata}
 shows how the magnitude of the
interaction varies. It can be seen from Tab.~\ref{tab:gamma_00mn} and Fig.~\ref{fig:relations_00mndata} that
$\left|\gamma_{00mn}\right|_{\text{nonzero}}$ reaches the maximum at $m=n$
for a fixed $m$ and decreases with ``the distance between $m$ and
$n$'' $|m-n|$. It also can be see that $\left|\gamma_{00mn}\right|_{\text{nonzero}}$
tend to zero when $n\rightarrow\infty$. Thus we have a conclusion
that a scatter process including one or two very high excited KK states
can be neglected.

\begin{figure*}[htbp]
    \makebox[\textwidth][c]{
  \includegraphics[width=0.4\textwidth]{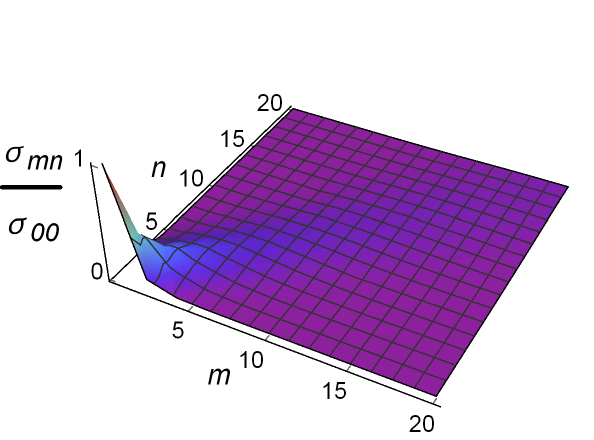}
   }
    \caption{{Illustration for the total cross section $\sigma_{mn}$ in (\ref{TotalCrossSection}) in unit of $\sigma_{00}$}.}
    \label{fig:TotalCrossSection_sigma_mn}
\end{figure*}

The cross section of the process $00\rightarrow mn$ is
\begin{eqnarray}
d\sigma_{mn}&=&(2\pi)^{4}\delta^{4}(p_{3}+p_{4}-p_{1}-p_{2})
     \frac{\left|\gamma_{00mn}\right|^{2}} {\left(-4p_{1}\cdot p_{2}\right)}\\ \nn
     &&*\frac{d^{3}\vec{p}_{3}} {(2\pi)^{3}2p_{3}^{0}}
     \frac{d^{3}\vec{p}_{4}} {(2\pi)^{3}2p_{4}^{0}},
\end{eqnarray}
where the subscripts $1$ and $2$ refer to the initial state particles
and $3$ and $4$ refer to the final state particles. Integrating
it we obtain the total cross section
\begin{equation}
\sigma_{mn}=\frac{1}{4\pi}\frac{\left|\gamma_{00mn}\right|^{2}}{\left(-p_{1}\cdot p_{2}\right)}\frac{{p^{2}}}{2(p_{1}^{0}+p_{2}^{0})^{2}},  \label{TotalCrossSection}
\end{equation}
where
\begin{eqnarray}
  {p^{4}} &=&\big(p_{1}^{0}+p_{2}^{0}\big)^{4}
        -2\big(p_{1}^{0}+p_{2}^{0}\big)^{2}m_{3}^{2}\\ \nn
        &-&2\big(p_{1}^{0}+p_{2}^{0}\big)^{2}m_{4}^{2}
        -2m_{3}^{2}m_{4}^{2}+m_{3}^{4}+m_{4}^{4},
\end{eqnarray}
it characterizes the size of the final state phase space. It can be
shown that the total cross section will decline with $m,n$, see Fig. \ref{fig:TotalCrossSection_sigma_mn} for illustration. For a
fixed center mass energy $E=p_{1}^{0}+p_{2}^{0}$ of the two zero
mode particles, the total cross section is zero when $m_{3}+m_{4} > E$.

Let us assume that we have very high initial energy so that a large
number of reaction branches are turned on. The branch ratio $R(00\rightarrow00)$
is always the largest one. When the initial energy $p_{1}^{0}+p_{2}^{0}\gg m_{3}$
and $p_{1}^{0}+p_{2}^{0}\gg m_{4}$, the variation of the size of the final state
phase space is negligible and the branch ratio $R(00\rightarrow mn)$
grows with $n$ and reaches its maximum at $n=m$, behaving just like
$\left|\gamma_{00mn}\right|$. On the other hand, larger value of
$n$ or $m$ suppresses the scattering amplitude drastically, as well
as suppresses the size of the final state phase space.

Therefore, the observable branch ratios $R(00\rightarrow mn)$ mainly
come from the ones with smaller $m$ and $n$. Furthermore, in an
inclusive process in which the final state involves a KK mode $n$,
the branch ratios of the processes that the final state includes another
KK mode close to $n$ are the most significant.

The ``${000n}$'' interaction is especially important because
it represents the amplitude that a KK particle decays to three zero mode
particles. We could deduce the decay rate $\Gamma_{n000}$ of a KK
particle of mode $n$ from $\gamma_{000n}$:
\begin{equation}
\Gamma_{n000}=\sum_{f}(2\pi)^{4}\delta(p_{f}^{0}-2k\sqrt{6n})\delta^{3}({\vec{p}}_{f})\lambda^{2}\left|\gamma{}_{000n}\right|^{2},
\end{equation}
where the subscript $f$ denotes the final state. The values of $\Gamma_{n000}$ in unit of $10^{-9}k^{3}$ are shown in Tab.~\ref{tab:varGamma_n000}. For the even modes $n$, the decay rate $\Gamma_{n000}$ decreases with mode number $n$.  For the odd modes, the decay channels $n\rightarrow000$ do not exist.
Instead, they have nonzero $\Gamma_{n100}$, see Tab.~\ref{tab:varGamma_n100}.

\begin{table*}[!htb]
\begin{center}
\begin{tabular}{|c|c|c|c|c|c|c|c|c|c|c|}
\hline
$n$ & 2 & 4 & 6 & 8 & 10 & 12 & 14 & 16 & 18 & 20\tabularnewline
\hline
$\Gamma_{n000}$ & 2.62 & 0.44 & 0.074 & 0.011 & 0.002 & 0.0003 & 0.00005 & 0.000008 & 0.000001 & 0.0000002\tabularnewline
\hline
\end{tabular}
\caption{The values of $\Gamma_{n000}$ in unit of $10^{-9}k^{3}$ for even $n$.}
\label{tab:varGamma_n000}
\end{center}
\end{table*}

\begin{table*}[!htb]
\begin{center}
\begin{tabular}{|c|c|c|c|c|c|c|c|c|c|}
\hline
$n$ & 3 & 5 & 7 & 9 & 11 & 13 & 15 & 17 & 19\tabularnewline
\hline
$\Gamma_{n100}$ & 1.88 & 1.26 & 0.45 & 0.02 & 0.004 & 0.0009 & 0.0002 & 0.00003 & 0.000006\tabularnewline
\hline
\end{tabular}
\caption{The values of $\Gamma_{n100}$ in unit of $10^{-9}k^{3}$ for odd $n$.}
\label{tab:varGamma_n100}
\end{center}
\end{table*}

The above two tables \ref{tab:varGamma_n000} and \ref{tab:varGamma_n100} show the same regular pattern: the decay rate vanishes when the mode number increases. The KK particle with higher modes seem
``isolate'' from the zero mode in the decay process.

\begin{figure*}[htbp]
    \makebox[\textwidth][c]{
    \includegraphics[width=0.3\textwidth]{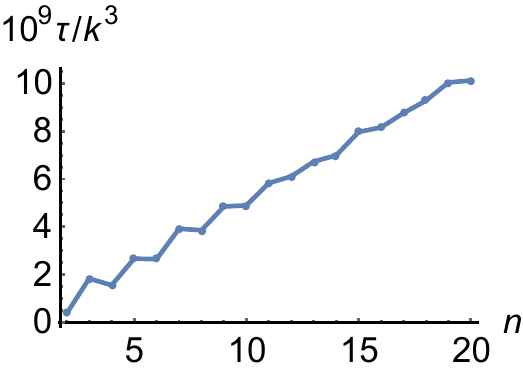}
    }
    \caption{The lifetimes of the KK particles.}
    \label{fig:lifetime_KKmodes}
\end{figure*}

For a KK particle with mode $n$, it could decay to three lower mode particles,
so there are various decay channels $n\rightarrow klm$. The lifetime
$\tau$ of the particle is the reciprocal of the sum of its decay
rates into all possible final states. Checking the lifetime of the KK
particles, we find that particles with higher mode have longer lifetime (see Fig. \ref{fig:lifetime_KKmodes}).
The KK particles with higher modes may be practically stable.

Note that the mass spectrum of the KK modes, the cross sections, the branch ratio, the decay rate and the lifetime of a KK particle are also affected by the fundamental coupling $\lambda$ in five dimensions and the new physics energy scale $k$ in the second model.

\section{Conclusions and Discussions} \label{sec_4}

In this paper we have investigated the effective action on the brane
from two five-dimensional braneworld models with a perturbation interaction
of a scalar field. We have discussed how a five-dimensional interaction
affects on our four-dimensional world, in the cases of PT and harmonic
potentials. We demonstrated some common features of these effects,
not just of these certain potentials. These conclusions could be applied to more general interactions other than $\phi^{4}$.

The first conclusion is that one or more KK particles will appear if we improve
the energy of particle collision, and new interactions between those
KK particles, including the zero mode particle, will arise at the
same time. The properties of these interactions can be seen by calculating
the effective coupling constants. Nevertheless, the new particles
corresponding to higher KK modes are more difficult to discover, not
only due to the requirement of higher energy, but also for the diminishing
of the involved effective interaction.

There is a obvious tendency of the effective coupling constants: $\gamma_{klmn}$
tends to vanish when $k,l,m,n$ ``depart'' from each other. Especially,
the zero mode $\varphi_{0}$ seems to decouple from other KK modes
$\varphi_{n}$ with large $n$.

With properties of KK particles, we might have an interesting observation.
The KK particles with higher modes have larger masse and longer lifetime,
and it almost does not interact with the ordinary matter in four-dimensional
world if its mode number $n$ is large enough. Exactly, this is one of the features of dark matter. Therefore, the KK particles with higher modes might be a candidate of dark matter if they are localized on the brane.

In the future work, we would like to study the interaction between
KK fermions and KK vectors from $\bar{\varPsi}\gamma^{M}A_{M}\varPsi$.
The method is analogous to what we performed in this paper. However,
there are significant differences between them. For a spinor field,
the left and right components of the zero mode can not been localized
on brane at the same time \cite{Bajc2000,Oda2000,Melfo2006,Volkas2007,Liu2008,Almeida2009,Liu1707.08541,LiLiu2017,Mazani2020}. If we regard the zero mode particle as
an electron, a contradiction would arise since it has been observed that electrons
have both chiralities. On the other hand, the four-dimensional effective electrodynamics
may not be gauge invariant, as a result of the massive KK modes of a vector field.

\section*{Acknowledgement}

This work is supported in part by the National Key Research and Development Program of
China (Grant No. 2020YFC2201503), the National Natural Science Foundation of China
(Grants No. 11875151, No. 12147175, and  No. 12247101), the 111 Project (Grant No. B20063) and Lanzhou City's scientific research funding subsidy to Lanzhou University.

\end{document}